\documentclass[aps,prd,
nofootinbib,
preprintnumbers,
twocolumn,
showpacs, 
floatfix,
superscriptaddress
]{revtex4}
\usepackage{amsmath}
\usepackage{amssymb}
\usepackage{epsfig}
\usepackage{graphicx}
\usepackage{multirow}
\usepackage{color}
\usepackage[normal]{subfigure}
\usepackage{rotating}
\usepackage{hyperref}

\begin{document}

\thispagestyle{empty}
\preprint{ULB-TH/12-20}

\title{Constraints on primordial black holes 
as dark matter candidates from star formation}

\author{Fabio Capela}
\email{fregocap@ulb.ac.be}
\affiliation{Service de Physique Th\'{e}orique, 
Universit\'{e} Libre de Bruxelles (ULB),\\CP225 Boulevard du 
Triomphe, B-1050 Bruxelles, Belgium}

\author{Maxim Pshirkov}
\email{pshirkov@prao.ru}
\affiliation{Sternberg Astronomical Institute, Lomonosov Moscow State University, 
Universitetsky prospekt 13, 119992, Moscow, Russia}

\affiliation{Pushchino Radio Astronomy Observatory, 
Astro Space Center, Lebedev Physical Institute  Russian Academy of Sciences,  
142290 Pushchino, Russia}

\affiliation{Institute for Nuclear Research of the Russian Academy of Sciences, 117312, 
Moscow, Russia}

\author{Peter Tinyakov}
\email{petr.tiniakov@ulb.ac.be}
\affiliation{Service de Physique Th\'{e}orique, 
Universit\'{e} Libre de Bruxelles (ULB),\\CP225 Boulevard du 
Triomphe, B-1050 Bruxelles, Belgium}

\begin{abstract} 
By considering adiabatic contraction of the dark matter (DM) during star
formation, we estimate the amount of DM trapped in stars at their birth.
 If the DM consists partly of
primordial black holes (PBHs), they will be trapped together with the rest of
the DM and will be finally inherited by a star compact remnant --- a white
dwarf (WD) or a neutron star (NS), which they will destroy in a short
time. Observations of WDs and NSs thus impose constraints on the abundance of
PBH. We show that the best constraints come from WDs and NSs in globular
clusters which exclude the DM consisting entirely of PBH in the mass range
$10^{16}{\rm g} - 3\times 10^{22}{\rm g}$, with the strongest constraint on the fraction
$\Omega_{\rm PBH} /\Omega_{\rm DM}\lesssim 10^{-2}$ being in 
the range of PBH masses $10^{17}{\rm g} - 10^{18}$~g.
\end{abstract}

\maketitle

\section{Introduction}
\label{sec:introduction}

Various observational evidence points at the existence of a new
matter component in the Universe, the
dark matter (DM) (for a recent review see, e.g.,
\cite{Bertone:2004pz,Bergstrom:2012fi}). 
Observations of the cosmic microwave background
imply that DM comprises $\sim 23$\% of the total energy budget of the
Universe, thus dominating in the matter sector, where the baryonic
component sums up to only 4\% \cite{Komatsu:2010fb}.
However, the nature of the DM remains
unknown, and masses of possible candidates range over many orders of
magnitude from a fraction of eV to many solar masses. Although most
popular candidates are new stable particles, other possibilities are
not excluded. 

In the early Universe density perturbations with high initial amplitude could
collapse forming black holes \cite{Hawking:1971ei}. If some of these black
holes survive until now they could constitute (at least) a fraction of the
DM. Properties of these primordial black holes (PBHs) make them a suitable DM
candidate: they are nonrelativistic and have subatomic size $r \sim
10^{-8}~\text{cm} \, (m_{\text{BH}}/10^{20} g)$, which makes them effectively
collisionless. Unlike most of the other DM candidates, PBHs do not require the
existence of new particle species.

The initial mass function of PBHs is flat in the case of a flat power spectrum
of primordial density fluctuations. However, models with strongly nonflat
mass function of PBHs can be constructed, see, e.g.,
Refs.~\cite{Dolgov:1992pu,Green:1999xm}.  The constraints at different
masses, therefore, should be considered independently.

Due to Hawking evaporation \cite{Hawking:1974rv}, the PBHs with masses $m_{\text{BH}}\leq 5\times 10^{14}~ \text{g}$
have lifetimes 
shorter than the present age of the Universe. Such PBHs thus cannot
contribute to the DM. 

PBHs with slightly larger masses emit $\gamma-$rays with energies
around $\sim 100$~MeV \cite{Page:1976wx}. Observations of the
extragalactic gamma-ray background with the Energetic Gamma Ray
Experiment Telescope (EGRET) \cite{Sreekumar:1997un} set an upper
limit on the cosmological density $\Omega_{\text{PBH}}$ of such PBHs
as a function of their mass, e.g. $\Omega_{\rm PBH}\leq 10^{-9}$ for
$m_{\text{BH}}= 10^{15}~\text{g}$ \cite{Carr:2009jm}. These
observations show that PBHs with masses $m_{\text{BH}}\leq 10^{16}
~\text{g}$ cannot constitute more than 1\% of DM. However, the 
constraints coming from the process of Hawking evaporation disappear for PBH
masses larger than $m_{\rm BH} \gtrsim 7\times 10^{16}$~g. 

The PBHs in the mass range $m_{\text{BH}}\lesssim 10^{19}-10^{20}~\text{g}$
can be constrained with the so-called ``femto lensing'' of the gamma-ray
bursts \cite{Gould:1992wz}. Present day observations of gamma-ray bursts
constrain the mass fraction of PBHs in the narrow mass range around
$m_{\text{BH}}\sim 10^{18}~\text{g}$ at several percent level
\cite{Barnacka:2012bm}.  The abundance of more massive PBHs can be constrained
from microlensing surveys. The EROS microlensing survey sets an upper limit of
8\% on the fraction of the Galactic halo mass in the form of PBHs with masses
in the range of $10^{26}~\text{g}<m_{\text{BH}}<3\times10^{34}~\text{g}$
\cite{Tisserand:2006zx}. At even higher mass scales,
$10^{33}~\text{g}<m_{\text{BH}}<10^{40}~\text{g}$, the analysis of the cosmic
microwave background can be used to constrain PBHs at the level of $10^{-7}$
\cite{Ricotti:2007au}.

The range of PBH masses from roughly $10^{17}$ to $10^{26}\text{g}$
remains essentially unconstrained, apart from the above-mentioned narrow
region around $10^{18}~\text{g}$. The aim of this paper is to constrain PBHs
as the DM candidates in this still allowed mass range. To this end, we
investigate the effect of PBHs on the evolution of compact stars -- neutron
stars (NSs) and white dwarfs (WDs).  The main idea is as follows. PBHs may be
captured by a star in the process of its formation. This leads to no
observational consequences until the evolution of the star reaches the stage
of a neutron star or a white dwarf. Then the accretion onto a PBH becomes
sufficiently fast to destroy the compact star in a short time
\cite{Kouvaris:2011fi,Kouvaris:2011gb,Kouvaris:2010jy}.  The region of PBH
masses and abundances where this happens with large probability is thus
excluded by observations of the existing neutron stars and white dwarfs.

The paper is organized as follows. In Sec. II we discuss the
gravitational capture of DM during the process of star formation. In
Sec. III we derive the constraints on the fraction of PBHs in the
DM from the existence of WDs and NSs. In Sec. IV we summarize the
results and present our conclusions. Throughout the paper, we use the
units $\hbar = c=1$.

\section{Capture of dark matter during star formation}
\label{sec:capture-dark-matter}
In this section we study the capture of dark matter during the star formation
process, neglecting all the DM interactions except the gravitational one. The
purpose is to estimate the total amount of DM captured inside newly formed
stars.

\subsection{Star formation stages}
\label{sec:star-form-stag}

Star formation occurs mainly in giant molecular clouds (GMCs). GMCs are dense
regions of the interstellar medium composed primarily of molecular hydrogen
($H_2$) with typical mass $M\sim\, 3\times 10^5 \,M_{\odot}$ and average
density $\rho \sim 550 \, \text{GeV cm}^{-3}$, which would imply a radius
of $17\text{pc}$ in the case of a spherical shape. A GMC is usually
composed of smaller overdense subclouds, i.e., clumps. In gravitationally
bound cores inside the clumps, individual stars are formed.

The formation of stars involves different stages. The first one corresponds to
the fragmentation of a GMC into gravitationally bound regions that are
initially supported against gravity by a combination of rotation and magnetic and
turbulent pressures \cite{1991ApJ...371..296M, 1987ARA&amp;A..25...23S}.  At
some point, as the cloud core loses its magnetic and turbulent support by
still poorly understood mechanisms like ambipolar diffusion
\cite{1987ARA&amp;A..25...23S}, the growing central concentration becomes
unstable to the gravitational collapse.  At the initial stage of the collapse,
the cloud has a uniform temperature, is rotating slowly and has an almost flat
density profile in the central part. 
At the end of this phase an opaque protostellar object in hydrostatic
equilibrium is formed at the center \cite{Larson69,Boss95,Bate:1998hg}. 

When the protostar is formed, it accretes from the surrounding disk increasing
its temperature. When the central object has accumulated most of its
main-sequence mass and the surrounding disk disperses, it is considered a
pre-main-sequence star. The main energy source for such an object is the
gravitational contraction, contrary to nuclear fusion for main sequence
stars. Therefore it is evolving on a Kelvin-Helmholtz timescale
$GM_{*}^2/(R_{*}L_{*})$, where $M_{*}, R_{*}$ and $L_{*}$ are the mass, the
radius and the luminosity of the pre-main-sequence star, respectively. 
This time is longer than the free-fall time $(R_*^3/GM_*)^{1/2}$.

\subsection{Adiabatic contraction}
\label{sec:adiab-contr}

The main mechanism of the capture of DM by stars at the time of formation is
the adiabatic contraction. Consider first this mechanism in general terms.

In this paper we will be  interested in systems that are dominated by
baryons. In this case the adiabatic contraction is easy to understand.  
When baryons contract losing energy by nongravitational
mechanisms, their time-dependent gravitational potential pulls the DM
particles along. The DM distribution thus develops a peak centered at the
baryon distribution.

If the change of the baryonic gravitational potential is slow, that
is, if the characteristic time of the baryonic contraction is much
larger than the free-fall time $t_{\rm ff}= (G\rho_0)^{-1/2}$, where
$\rho_0$ is the baryonic density of a cloud, the DM distribution is
determined by the (approximate) conservation of the adiabatic
invariant
\begin{equation}
\oint p dq = E T
\label{eq:invariant}
\end{equation}
where $p$ and $q$ are the phase-space coordinates of a DM particle of
energy $E$ and orbital period $T$. Moreover, the angular momentum is
conserved for each DM particle as long as the potential is
central. From these conserved quantities, a relation between the
initial orbital radius and the final one can be derived
\cite{Blumenthal:1985qy,Gnedin:2004cx,Sellwood:2005pq,1999A&amp;A...343....1D}.

Regardless of whether the contraction of DM is adiabatic or not, the
phase-space density of DM has to be preserved all along the contraction
process, as dictated by Liouville's theorem. For an initial 
Maxwellian velocity
distribution of DM with the dispersion $\bar{v}$, the maximum phase-space 
density is at zero velocity and equals \cite{Tremaine:1979we} 
\begin{equation}
\mathcal{Q}_\text{max}=
\left(\frac{3}{2\pi}\right)^{3/2}\frac{\bar{\rho}_\text{DM}}{m_\text{DM}^4\bar{v}^3}
\end{equation}
where $\bar{\rho}_\text{DM}$ is the space density of DM, and $m_\text{DM}$ is
the DM mass.  The effect of the
adiabatic contraction is to fill all the allowed phase-space with the density
close to the maximum value.

In the case of circular orbits the conservation of the angular momentum and
the adiabatic invariant (\ref{eq:invariant}) implies the conservation of the
quantity $rM(r)$, where $M(r)$ is the mass within the radius $r$. Suppose a
baryonic cloud, which was initially a uniform sphere of radius $\bar R$,
contracts to a compact object of a negligible size.  Assuming that the DM
particles move on circular orbits, the initial uniform DM density $\bar
\rho_{\rm DM}$ is modified as follows \cite{1978AJ.....83.1050S}:
\begin{equation}
\rho_{\rm DM} (r) = {1\over 4} \bar \rho_{\rm DM}  \left({\bar R \over r}\right)^{9/4},
\label{powlaw}
\end{equation} 
provided the adiabatic approximation holds. 

In realistic cases the orbits of DM particles are not circular. The question
thus arises whether Eq.~(\ref{powlaw}) is a good approximation in realistic
situations. An exact calculation has been performed in another limiting case
of purely radial orbits \cite{Spolyar:2007qv,Bambi:2008uc}. The results were
found to be roughly compatible with the case of circular orbits. However, this
is not a realistic case either. Another question of practical importance is
the domain of validity of the adiabatic approximation. Formally, it requires
the time of collapse $t_c$ to be much longer that the free-fall time $t_{\rm
  ff}$. In Ref.~\cite{Jesseit:2002tj} high-resolution numerical simulations
have been performed and it has been shown that the adiabatic contraction may
remain a good approximation, depending on the potential, even when $t_c\simeq
t_{\rm ff}$.

To clarify these issues, we have performed the following 
numerical simulation. For the baryonic distribution that is responsible for
the time-varying external gravitational potential we took the sum of a uniform
spherical cloud and a point mass in the center. The point mass was zero at the
initial moment of time and then increased linearly with time, while the mass
of the spherical part, always uniform in density, decreased in such a way that
its sum with the point mass remained constant. The time $t_c$ over which all
the mass was transferred from the cloud to the central object was treated as a
free parameter.

The DM particles were injected at $t=0$ with an initial uniform distribution
in position and velocity. The initial density profile was taken to be constant
over the volume of the cloud to mimic the physical properties of prestellar
cores. The particles were injected one by one, which corresponds to neglecting
the DM contribution to the gravitational potential. Those particles that have
positive total energy at $t=0$ (and thus are not gravitationally bound to the
system) were discarded; the remaining ones were evolved numerically in the
time-dependent gravitational potential. 
At a random time $t>t_c$, the positions
of these particles were sampled in order to reconstruct, after many simulations, the
final density profile.  As a consistency test, we have also performed an
identical simulation with initial velocities of DM particles generated in such
a way that the particles move on circular orbits.
\begin{figure}
\includegraphics[width=0.9\columnwidth]{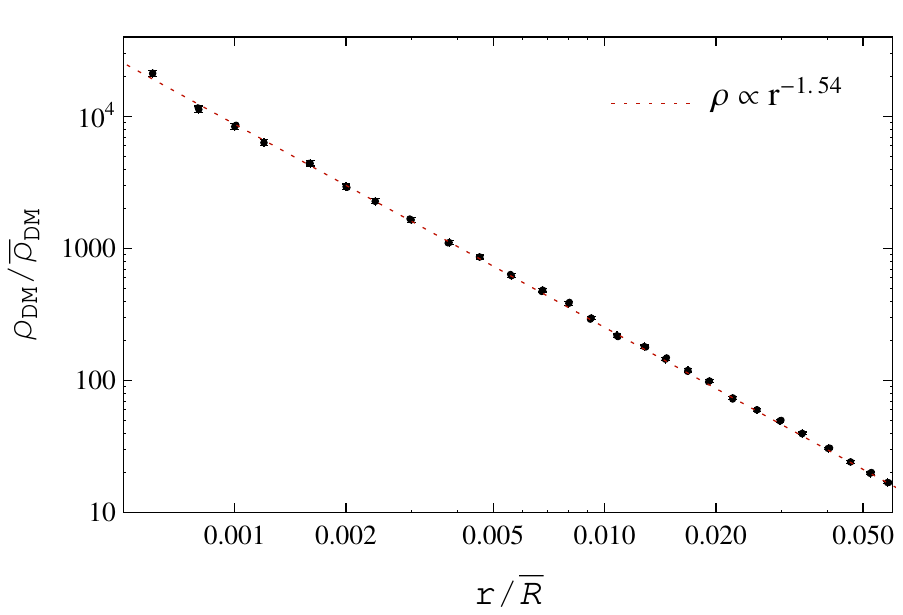}
\caption{\label{fig:sim} DM density profile obtained from the simulation
after the adiabatic formation of a star.
The inner part of the profile has the slope close to $-3/2$, as expected from Liouville's theorem for the uniform initial velocity distribution.}
\end{figure} 

In the case of circular orbits we have reproduced the density profile
\eqref{powlaw}, even for a relatively rapid change of the external potential,
$t_c= 3~t_{\rm ff}$.  In the case of random initial velocities, however, the
inner profile was found to have a slope close to $-3/2$, 
\begin{equation}
\rho_{\rm DM} (r) = {1\over 2} \bar \rho_{\rm DM} 
\left({\bar R \over r}\right)^{3/2}
\label{eq:profile-random}
\end{equation}
as represented in Fig.~\ref{fig:sim}.  These results are in agreement with
Liouville's theorem, since the final DM velocity goes as $v(r)\propto
r^{-1/2}$.  Since random initial velocities appear to be a better
approximation to realistic initial conditions than the circular ones, and
because the profile (\ref{eq:profile-random}) gives more conservative
estimates, in the rest of this paper we use the profile
(\ref{eq:profile-random}).

\subsection{DM bound to a baryonic cloud}
\label{sec:dm-density-baryonic}

As is clear from the above discussion, only DM gravitationally bound to a
baryonic cloud is subject to the adiabatic contraction when the cloud
collapses. Therefore, to set the initial conditions for the adiabatic
contraction we need to estimate the amount of DM that is gravitationally bound
to a cloud. 

We will assume that the matter density is dominated by baryons, as is the case
in the star-forming regions. 
When
the overdensity of baryons is formed, some amount of DM ends up gravitationally
bound to the baryonic cloud. Consider the formation of a spherical cloud of
radius $R_0$ and baryonic density $\rho_0$.  Our goal is to estimate the
density of DM bound to the cloud, $\rho_{\rm DM,\, bound}$, given the mean
density of DM, $\bar \rho_{\rm DM}$. We will assume that originally the DM
particles have the Maxwellian distribution in velocities with the dispersion
$\bar v$,
\begin{equation}
d n = \bar n_{\rm DM} \left({3\over 2\pi \bar v^2}\right)^{3/2}
\exp \left\{{-3v^2\over 2 \bar v^2}\right\} d^3v , 
\label{eq:Maxwell}
\end{equation}
where $\bar n_{\rm DM} = \bar \rho_{\rm DM}/m$, with $m$ being the mass of the DM
particle. We will see that in the cases of interest the velocities of bound
particles are much smaller than $\bar v$, and thus the precise shape of the
distribution is not essential.

After the formation of a baryonic cloud, the gravitational potential felt by
DM particles becomes of order
\[
\phi\sim \phi_0 = 2\pi G\rho_0 R_0^2. 
\]
Those particles with kinetic energies smaller than $\phi_0$ (equivalently,
velocities $v<v_0=\sqrt{2\phi_0}$) become gravitationally bound. Their number
density is obtained by integrating Eq.~(\ref{eq:Maxwell}) up to
$v=v_0$. Multiplying by the DM mass, one has
\begin{equation}
\rho_{\rm DM,\,bound} = \bar \rho_{\rm DM}{4\pi\over 3} 
\left({3 |\phi_0| \over \pi \bar v^2}\right)^{3/2}
\label{eq:nbound-phi}
\end{equation}
\begin{equation}
= \bar \rho_{\rm DM}{4\pi\over 3} 
\left({6 G \rho_0 R_0^2\over \bar v^2}\right)^{3/2},
\label{eq:nbound-slow}
\end{equation}
where we have assumed $v_0\ll \bar v$ and thus set the exponential to 1. 

\subsection{Globular clusters}
\label{sec:globular-clusters} 
\begin{table}
\begin{tabular}{l|c|l}
$M_*/M_\odot$ & $\rho_0$, GeV\,cm$^{-3}$  & $R_0$, AU \\\hline
1 & $10^6$ & $4300$ \\ 
 2 &$1.8\times 10^6$ & $4450$ \\ 
3 &$2.4\times 10^6$ & $4620$ \\ 
4 &$3.1\times 10^6$ & $4710$ \\ 
5 &$3.6\times 10^6$ & $4780$ \\ 
6 &$4.2\times 10^6$ & $4840$ \\ 
7 &$4.8\times 10^6$ & $4880$ \\ 
8 &$5.3\times 10^6$ & $4930$ \\ 
10 &$6.4\times 10^6$ & $5000$ \\ 
12 &$7.4\times 10^6$ & $5060$ \\ 
15 & $8.8\times 10^6$ & $5130$ 
\end{tabular}
\caption{\label{tab:prestellar-cores} The parameters of prestellar cores
  used in the estimates. }
\end{table}

Globular clusters (GCs) are gravitationally bound systems consisting of $10^4$
to $10^7$ stars with average diameters ranging from 20 to 100 pc. There are
about 100 GCs known in our Galaxy.  A typical GC has a
baryonic mass of $\mbox{(a few)} \times 10^5 M_{\odot}$ and a core radius of
$1-2, \,\text{pc}$. The age of GCs is about $8$ to $13.5$~Gyr \cite{Dotter2010}, and as such they
are the oldest surviving stellar subsystems in the Galaxy, made up of the
population II stars, white dwarfs, neutron stars and black holes.

There are two classes of scenarios for GC formation. According to the
primordial, or ``DM-dominated'' one, GCs were formed by the infall of baryonic
matter into the gravitational wells of the DM density peaks at redshifts
$z>10$ \citep{Peebles1984,Bromm2002,MS2005a,Moore2006, Boley2009,
  Griffen2010}. The second is the ``baryon-dominated'' scenario, according to
which GCs were formed in predominantly baryonic processes like major mergers,
hydrodynamical shocks and so on, mostly in protogalaxies that later assembled
into the Milky Way \citep{Fall1985,Ashman1992,Kravtsov2005,
  Muratov2010}. Moreover, both mechanisms could be at work because the
observed distribution of metallicity in GCs is clearly bimodal, so that
metal-poor GC could be of cosmological origin, while metal-rich GCs could be
formed in the course of mergers \citep{Brodie2006}. Although there is no
evidence of DM presence in the GCs now \citep{Moore1996}, it was shown that it
could be present at the formation time and subsequently tidally stripped due
to interactions with the host galaxy \citep{MS2005b}. We will assume in what
follows that at least some of the GCs resided in DM minihaloes in the past,
and concentrate on those.

The DM density in the central regions of GCs has been estimated in
Ref.~\cite{Bertone:2007ae} by making use of the formalism developed in
Refs.~\cite{MS2005a,MS2005b}.  The conclusion was that the present-day
DM density close to the core of a GC is of order $\rho_{DM}\sim
2\times10^3~\text{GeV}~\text{cm}^{-3}$, the estimate being rather
insensitive to the original halo mass. This result is in concordance
with the N-body simulations \cite{MS2005a,MS2005b} suggesting that the
inner part of DM halos survives successive tidal interactions with the
host galaxy. As has been stressed in Ref.~\cite{Bertone:2007ae}, the
number cited includes the effect of dynamical heating of DM by the
stars comprising the cluster, which reduces the DM density. In our
estimates this effect is irrelevant since we are interested in the
evolution stage prior to the star formation. With no heating, the DM
density in the core would be $\rho_{DM}\sim
10^4~\text{GeV}~\text{cm}^{-3}$, which we adopt in what follows.

Another important parameter is the value of the DM velocity dispersion in
GCs. As stars in the GC are collisionless and behave essentially as DM
particles, this parameter can be extracted directly from
observations. Although there is quite a bit of scatter, typical observed GCs
have the velocity dispersion around $\bar{v}=7~\text{km}\,{\rm s}^{-1}$
\cite{GC-velocity-dispersion}.  

Let us now turn to the prestellar cores. Their typical parameters are known
from observations carried out with the SCUBA instrument \cite{Kirk:2005ng}. The data set is well fitted by the
Bonnor-Ebert profile \cite{Kirk:2005ng} which, for our purposes, can be
approximated by the flat core of radius $R_0$, containing the baryonic mass
$M_*$ of the future star.

Two cases will be of interest in what follows: stars with masses
$ 1M_\odot \leq M_*\leq 7 M_\odot$ which are typical progenitors of a WD and supermassive stars of
masses $M_*\geq8 M_\odot$ progenitors of NSs.
In all cases the gravitational potential of the prestellar core is (much)
smaller than that of the GMC, so one may use again Eq.~(\ref{eq:nbound-phi})
to estimate the density of DM gravitationally bound to the core. We use the
parameters of prestellar cores that are listed in
Table~\ref{tab:prestellar-cores}. As has been already mentioned, the
formation of the prestellar cores relies on the nongravitational energy loss
mechanisms and thus is expected to be slower than the free fall.

Making use of Eq.~(\ref{eq:profile-random}),  one obtains the total DM mass contained in
a star formed within the GCs as listed in Table~\ref{tab:GC-masses}. 
These values were calculated with the DM density and velocity dispersion given
above; for different values of these parameters the mass of the bound DM
should be rescaled by the factor
\begin{equation}
\left({\bar v \over 7 \,{\rm km/s}}\right)^{-3}
\left({\bar \rho_{\rm DM} \over 10^{4} \,{\rm GeV/cm}^3}\right)
\label{eq:factor}
\end{equation}
which may be different for different GCs. 
\begin{table}
\begin{tabular}{l|c|l}
$M_*/M_\odot$ &  $\rho_{\rm PSC}$, GeV\,cm$^{-3}$   & $M_{\rm bound}, \,{\rm g}$ \\\hline
1  & $2\times 10^1$  & $4.4\times 10^{19}$\\
2  & $5.2\times 10^1$  & $2.5\times 10^{20}$\\ 
3  & $9.2\times 10^1$  & $7.2\times 10^{20}$\\ 
4  & $1.4\times 10^2$  & $1.5\times 10^{21}$\\
5  & $1.9\times 10^2$  & $2.6\times 10^{21}$\\
6  & $2.4\times 10^2$  & $4.2\times 10^{21}$\\
7  & $3\times 10^2$  & $6.2\times 10^{21}$\\
8  & $3.6\times 10^2$  & $8.7\times 10^{21}$\\
10 & $5\times 10^2$  & $1.6\times 10^{22}$\\ 
12  & $6.4\times 10^2$  & $2.4\times 10^{22}$\\
15 & $8.7\times 10^2$  & $4.3\times 10^{22}$
\end{tabular}
\caption{\label{tab:GC-masses} Density of DM bound to the prestellar core,
  $\rho_{\rm PSC}$, and the total mass $M_{\rm bound}$ of  DM
  contained in a star right after its formation in a GC with the
  central DM density $\rho_{DM}\sim 10^4~\text{GeV}~\text{cm}^{-3}$
  and velocity dispersion $\bar{v}=7~\text{km}\,{\rm s}^{-1}$ for
  different star masses. }
\end{table}

\section{Constraints on primordial black holes}
\label{sec:constraints}

So far the discussion has been general and does not depend on the DM
nature. Consider now specifically the case of primordial black holes. The PBHs
that end up inside a star when the latter is formed start accreting and
gravitationally pulling on the surrounding matter, lose their momentum ,and
gradually sink to the center. The sinking process is slow, so that the
characteristic time may exceed the age of the star. Because of their slow
accretion and small number, the presence of the BH has no observable effect on
the star's evolution at this stage.

When a star polluted by PBHs evolves into a compact object (WD or NS), some of
the BHs get inside the compact remnant. Because of a much higher density, the
accretion is now more efficient and the PBHs, if present inside the remnant,
rapidly consume the latter. The observation of WDs and NSs thus implies
constraints on the abundance of PBHs, which has to be such that the
probability to get a PBH inside NS or WD is much less than 1. 

To quantify this statement, we calculate the number $N_{\rm BH}$ of BHs that
would sink down to the future radius $r_f$ of the compact remnant by the end
of the star evolution and, thus, would end up inside the star remnant. If
$N_{\rm BH} <1$, no constraints arise. If $N_{\rm BH} >1$, the maximum allowed
fraction of BHs in the total amount of DM is
\begin{equation}
{\Omega_{\rm PBH}\over \Omega_{\rm DM}} \leq {1\over N_{\rm BH}}.
\label{eq:omegaBH}
\end{equation}
Thus, in the range of PBH masses where $N_{\rm BH} >1$, PBHs cannot constitute
all of the DM.

The PBHs that are eventually trapped by the compact remnant initially occupy
some spherical volume of radius $r_c$, which we call the ``collection region''.
Knowing $r_c$ as a function of the PBH mass $m_{\rm BH}$ and the DM
distribution inside the star at the time of formation allows one to calculate
$ N_{\rm BH}$ as follows,
\begin{equation}
N_{\rm BH} =
M_{\rm DM}(r_c)/m_{\rm BH}, 
\label{eq:Nbh}
\end{equation}
where $M_{\rm DM}(r)$ is the DM mass contained in the radius $r$ at the time
of the star formation. 

The sinking of the PBH inside the star has been considered in
Ref.~\cite{Bambi:2008uc}. The dynamical friction force per unit PBH mass is
given by the Eq.~(16) of Ref.~\cite{Bambi:2008uc}. Multiplied by the PBH
velocity, this gives the PBH energy loss rate $dE/dt$. On the other hand,
assuming circular orbits, $dE/dt$ can be expressed in terms of the change of
the orbit radius. Equating the two gives a closed first-order differential
equation for the orbit radius as a function of time, $r(t)$. We derive and
solve the corresponding equation numerically in
Appendix~\ref{sec:sinking-bh-center}, assuming the star model with the
polytrope index $n=3$.

\begin{figure}
\begin{picture}(220,150)
\put(0,0){\includegraphics[width=0.9\columnwidth]{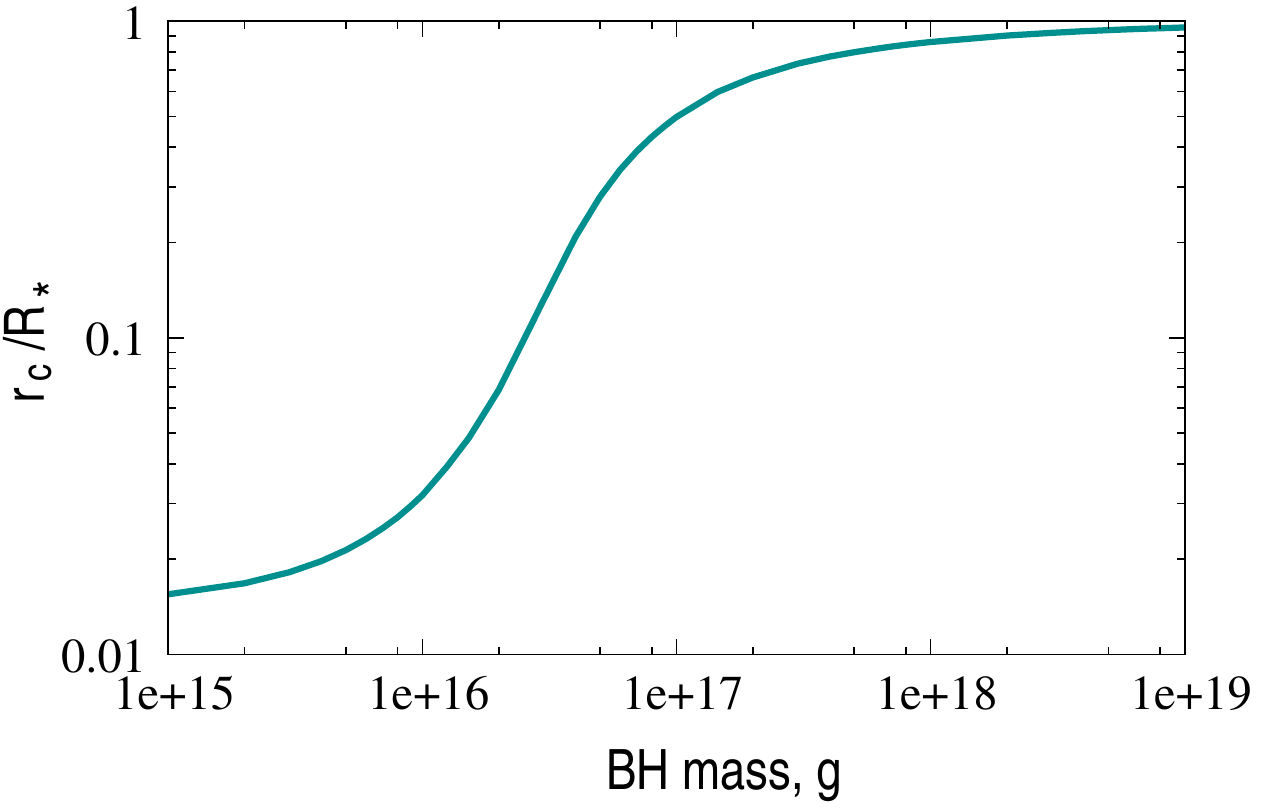}}
\end{picture}
\caption{\label{fig:r_i} The dependence of the size $r_c$ of the collection
  region (the region from which the PBHs captured by the star at its
  formation have enough time to sink to within the radius of the future
  compact remnant, WD or NS) on the PBH mass, corresponding
   to the case of WD for $M_*=M_\odot$. }
\end{figure}

Having found the dependence $r(t)$ for a given BH mass, we fix the final
radius $r_f = r(t_*)$ to be the size of the compact object (NS or WD). Here
$t_*$ is the lifetime of the star. We then determine the collection radius as
$r_c = r(0)$. (In practice, it is more convenient to run the evolution
equations backwards in time starting from $r=r_f$.) The dependence of $r_c$ on
the BH mass is shown in Fig.~\ref{fig:r_i} for 
$M_*=M_\odot$ and $r_f =r_{\rm WD} = 10^4\,{\rm km}$.

At small BH masses the dynamical friction is inefficient and the
collection radius $r_c$ is not very different from $r_f$. As the BH mass gets
larger the friction becomes more efficient, so that $r_c$ grows and eventually
becomes close to the star radius. The transition is quite rapid; the value of
$m_{\rm BH}= m_{\rm trans}$ at which it occurs can be understood analytically
from the behavior of Eq.~(\ref{eq:energy-loss-eq}); see
Appendix~\ref{sec:sinking-bh-center} for details. It corresponds to the smallest
$m_{\rm BH}$ for which the collection radius is still close to the star size
(i.e., the lifetime of the star is still sufficient for a BH to sink to
$r=r_f$ starting near the surface).  By an order of magnitude, it is given by
\[
m_{\rm trans} \sim 4\times 10^{17} \,{\rm g} 
\]
\begin{equation}
\times \left({t_*\over 10~{\rm Gyr}}\right)^{-1}
\left({M_*\over M_\odot}\right)^{1/2}
\left({R_*\over R_\odot}\right)^{3/2}.
\label{eq:mtrans}
\end{equation}
Note that there is no dependence on $r_f$ because the final stages of the BH
sinking are exponential, and these are the (longer) initial stages that
set the overall time scale. 

The number of BHs inside the collection region $N_{\rm BH}$ can be found from
the total DM mass trapped by the star (see Table~\ref{tab:GC-masses}) and the
DM density profile inside the star. The DM profile inside the star after the
adiabatic contraction is determined by Eqs.~(\ref{powlaw}) and
(\ref{eq:profile-random}) and the baryonic density, for which we assume the
density profile of the polytrope $n=3$ model. From
Eq.~(\ref{eq:profile-random}), the DM and baryonic masses
are related as follows,
\begin{equation}
M_{\rm DM} (r) 
= M_{\rm bound}  \left({M(r) r^3 \over M_* R_*^3} \right)^{1/2},
\label{eq:Nbh2}
\end{equation}
where $M(r)$ and $M_{\rm DM}(r)$ are the baryonic and DM mass within the radius
$r$, respectively.  The number of BH within $r_c$ is then given by
Eq.~(\ref{eq:Nbh}).

\begin{figure}
\begin{picture}(220,150)
\put(-10,0){\includegraphics[width=0.95\columnwidth]{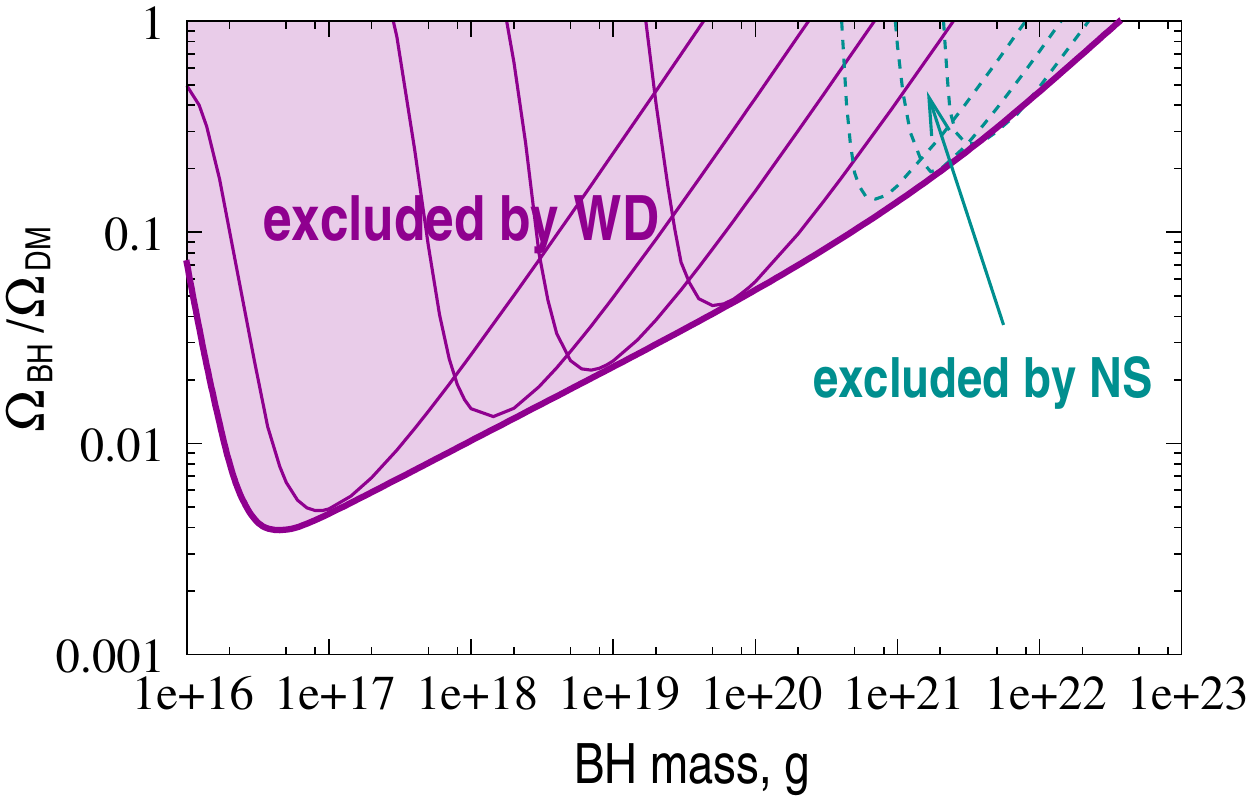}}
\end{picture}
\caption{\label{fig:constraints}
Constraints on the fraction $\Omega_{\rm
  PBH}/\Omega_{\rm DM}$. Purple shaded 
region is excluded by observations of WDs and NSs in the centers of globular
clusters. Thin curves show the exclusions from different star masses. }
\end{figure}

The resulting constraints on the fraction of PBHs in the total amount of DM
are shown in Fig.~\ref{fig:constraints}. Purple shading shows the region
excluded by the observations of WDs and NSs in the globular clusters. Thin
curves show the exclusion regions resulting from different star masses. One can
see that the constraints from WDs and NSs complement each other and together
cover the range of masses from $10^{16}$~g to $3\times 10^{22}$~g.

The shape of the excluded regions is similar in all cases shown in
Fig.~\ref{fig:constraints}. It can be understood from the mass dependence of
the collection radius $r_c$, Fig.~\ref{fig:r_i}, as follows. At large PBH
masses, the size of the collection region is close to the star size, so that
$M_{\rm DM}(r_c)\simeq M_{\rm bound}$ and the maximum PBH fraction $\Omega_{\rm
  PBH}/\Omega_{\rm DM}$ scales like $m_{\rm BH}$, i.e., the constraints
improve at smaller masses. However, at some point around $m_{\rm BH} \sim m_{\rm
  trans}$ the collection radius $r_c$ decreases
(cf. Fig.~\ref{fig:r_i}) and the constraints relax very rapidly.

\section{Conclusions} 
\label{sec:conclusions}

We have derived the constraints on the abundance of PBH from observations of
the existing WDs and NSs. The origin of these constraints is as follows. If
PBHs were present at the time of star formation, i.e., at $z\lesssim 10$, they
would pollute the newly formed stars and, after sinking to the center, would
end up in the compact remnant resulting from the star evolution (WD or NS).
Once inside the remnant, PBHs would rapidly destroy it by accretion. Mere
observations of WDs and NSs, therefore, impose constraints on the abundance of
PBHs.

We have found that the most stringent constraints come from
observations of WDs and NSs in globular clusters. WDs and NSs
are sensitive to the mass ranges $10^{16}\,{\rm g} \lesssim m_{\rm BH}
\lesssim 10^{21}\,{\rm g}$ and $10^{21}\,{\rm g} \lesssim m_{\rm BH}
\lesssim 3\times 10^{22}\,{\rm g}$, respectively, thus complementing each
other. Everywhere in this mass range the PBHs are excluded as
comprising all of the DM. The best constraint on the PBH fraction
$\Omega_{\rm PBH}/\Omega_{\rm DM}\lesssim 10^{-2}$ was found for
$m_{\rm BH}$ in the range $10^{17}\,{\rm g} - 10^{18}\,{\rm g}$.

The constraints derived from the globular clusters are based on the assumption
that at least some of those were formed in a primordial DM-dominated
environment. As a word of warning, it should be noted that this issue is still
debated in the literature. For instance, observations of a low-metallicity
cluster NGC 2419 \cite{Baumgardt2009,Conroy:2010bs} seem to indicate that its
mass-to-light ratio is in a good agreement with what is expected for a pure
baryonic system.  However, NGC 2419 has a number of extreme
properties \cite{Mackey2005,Cohen2012,Cohen2010} that make the globular
cluster nature of this object questionable.  In addition, high-resolution
N-body simulations \cite{MS2005a,MS2005b} indicate that the mass-to-light
ratio may not be sensitive to the presence of the DM component in GCs.

In order to derive the constraints on the PBH abundance we have investigated
the baryonic contraction of the DM during the star formation process. In
particular, we have calculated numerically the resulting DM profile and found
the slope close to $-3/2$. We also
estimated the total amount of DM that is trapped inside the star at the time of
its formation. This part of our results is not specific to any particular form
of the DM.

\acknowledgments

The authors are indebted to M.~Fairbairn, M.~Gustafsson and M.~Tytgat for
valuable discussions and comments, and to S.~Sivertsson for pointing out an inconsistency in the first version of this paper. The work of F.C. and
P.T. is supported in part by the IISN and the Belgian Science Policy (IAP
VI-11). The work of P.T. is supported in part by the ARC project Beyond
Einstein: Fundamental Aspects of Gravitational Interactions and by the Russian
Federation Ministry of Education under Contract No. 14.740.11.0890. The work of
M.P. is supported by RFBR Grant No. 12-02-31776 mol\_a.

\appendix %
\section{Sinking of BH in the star}
\label{sec:sinking-bh-center}
Here we consider the energy loss by a BH that is orbiting inside a star
gradually sinking to the center. The
star is assumed to have an $r$-dependent density $\rho(r)$ and temperature
$T(r)$ and the mass $M(r)$ enclosed inside the radius $r$. Our purpose is to
derive the equation that governs the evolution of the BH orbit due to the
dynamical friction, assuming the orbit is circular and changes slowly. The
question of dynamical friction was considered in a general context , e.g., 
in Refs.\cite{1949RvMP...21..383C,1987gady.book.....B}. 

The BH moving through a star experiences a dynamical friction force
\cite{1987gady.book.....B} that can be written as follows, 
\begin{equation}
{{\bf f}\over m_{\rm BH}} = - \gamma(v) {\bf v} , 
\label{eq:friction_force}
\end{equation}
where 
\[
\gamma(v) = 4 \pi G^2 \rho(r) m_{\rm BH} \ln(\Lambda) {F(X)\over v^3},
\]
\begin{equation}
F(X) = {\rm erf}(X) - 2X\exp(-X^2)/\sqrt{\pi},
\label{eq:FofX}
\end{equation}
\[
X = v/(\sqrt{2}\sigma),
\]
$\rho(r)$ is the density of the baryonic gas comprising the star, $\sigma$ is
the velocity dispersion of the particles $\sigma~=~\sqrt{T/m}$ with
temperature $T$ and mean molecular weight $m$ ($m\simeq 1.6$~GeV for a main sequence
star), and $\ln(\Lambda) \simeq \ln (M_*/m_{\rm BH})\simeq 30$ is the Coulomb
logarithm \cite{1987gady.book.....B}. Multiplying Eq.~(\ref{eq:friction_force})
by ${\bf v}$ gives the total BH energy loss rate per unit mass, $dE/dt=-
\gamma(v)v^2$.

Making use of the relation 
\begin{equation}
v^2 = GM(r)/r,
\label{eq:velocity}
\end{equation} 
the same energy loss rate can be expressed through the change of the 
radius $r$ of the BH orbit, 
\begin{equation}
{dE\over dt} = {d\over dt} \left({1\over 2} v^2 + U(r) \right)
= {dr\over dt} {v^2\over 2r} 
\left\{ {4\pi r^3\over M(r)} \rho(r) + 1 \right\},
\label{eq:dE/dt}
\end{equation}
where $U(r)$ is the gravitational potential.  Equating the two quantities and
simplifying by $v^2$ one gets
\begin{equation}
{dr\over dt} {1 \over 2r} 
\left\{ {4\pi r^3\over M(r)} \rho(r) + 1 \right\} = - \gamma(v) . 
\label{eq:dE/dt-2}
\end{equation}
By virtue of Eq.~(\ref{eq:velocity}), this is a closed differential
equation for the BH orbit radius $r(t)$ as a function of time. Note that this
equation is not equivalent to Eq.~(24) of Ref.~\cite{Bambi:2008uc} because the
contribution of the gravitational potential (the term $U(r)$ in
Eq.~(\ref{eq:dE/dt})) has been missed there.

Let us rewrite this equation in the form convenient for the numerical
solution. Define the dimensionless quantities
\begin{eqnarray}
\nonumber
x & = & r/R_*,\\
\nonumber
\tau & = & t/t_0,\\
\nonumber
\tilde \rho(r) & = & \rho(r)/\rho(0),\\
\nonumber
\tilde{M}(r)  & = & M(r)/M_*,\\
\nonumber
\tilde{T}(r)  & = & T(r)/T(0),
\end{eqnarray}
where 
\begin{equation}
t_0 = {M_*^{3/2} \over 2\pi \sqrt{G} \rho(0) m_{\rm BH} R_*^{3/2} \ln\Lambda}
\label{eq:t0}
\end{equation}
\begin{equation}
\simeq 4.2\times 10^3\, {\rm yr} 
\left({m_{\rm BH}\over 10^{22}{\rm g}}\right)^{-1}
\left({M_*\over M_\odot}\right)^{1/2}
\left({R_*\over R_\odot}\right)^{3/2}. 
\label{eq:scaling}
\end{equation}
The profiles $\tilde \rho(r)$ and $\tilde{T}(r)$ are determined by the star model. 
The normalization parameters $\rho(0)$, $T(0)$, $M_*$,
and $R_*$ are not independent. They obey the following two relations, 
\begin{eqnarray}
\nonumber
{GM_*m\over R_* T(0)} &=& 1.17\\
\nonumber
{\rho(0) R_*^3 \over M_*} &=& 12.9,
\end{eqnarray}
where, as before, $m$ is the mean molecular weight.  The scaling in
Eq.~(\ref{eq:scaling}) takes into account these relations.

In terms of the dimensionless quantities, Eq.~(\ref{eq:dE/dt-2}) becomes
\begin{equation}
{dx\over d\tau} = - {x^{5/2} \tilde\rho(x) \over f(x) \tilde M^{3/2}(x)}
F(X).
\label{eq:energy-loss-eq}
\end{equation}
Here we have introduced the function 
\[
f(x) = {1\over 4} \left\{ 4\pi {r^3\rho(r) \over M(r)} + 1\right\}
= {1\over 4} \left\{ 163 {x^3\tilde \rho(x) \over \tilde M(x)} + 1\right\}
\]
which varies between 1 in the star center $x=0$ and $1/4$ at the star surface
$x=1$. The variable $X$ is in turn a function of $x$ which can be expressed as
follows,
\[
X = \left({GmM(r)\over 2r T(r)}\right)^{1/2} = 0.765 
\left({\tilde M(x) \over x \tilde T(x)} \right)^{1/2},
\]
while the function $F(X)$ is defined in Eq.~(\ref{eq:FofX}). At small values
of $X$, this function behaves as $4X^3/(3\sqrt{\pi})$; at large $X$ it
asymptotes to 1. 

A useful analytical insight into the behavior of Eq.~(\ref{eq:energy-loss-eq})
can be obtained by considering two limiting cases. At small values of $x$ such
that the parameters of the star can still be approximated by their core values
one has
\[
X \simeq 5.62 x.
\]
At small $x$ such that $X\ll 1$ Eq.~(\ref{eq:energy-loss-eq}) becomes 
\[
{dx\over d\tau} = - 0.337 x,
\]
whose solution is $x(t) = \exp(- 0.337t/t_0)$ with $t_0$ given by
Eq.~(\ref{eq:t0}). In this regime, valid for the final approach by a sinking BH
of the radius $r_f$ (the radius of a future compact object), the
characteristic time scale is 
\[
(\Delta t)_2 \simeq 3\ln(r_0/r_f)\,t_0,
\]
where $r_0$ is some initial radius. 

At moderately small $x$, such that $X\gtrsim 1$, Eq.~(\ref{eq:t0}) takes the
form
\[
{dx\over d\tau} = - {1\over 397\, x^2},
\]
which gives the evolution time $(\Delta t)_1$ from $x_1$ to $x_2$, 
\[
(\Delta t)_1 = 132\, t_0 (x_1^3-x_2^3) \simeq 10^2 \times t_0,  
\]
where we have set $x_1^3\sim 1$ and neglected $x_2^3$. We see that this first
stage is typically longer than the second, $(\Delta t)_1>(\Delta
t)_2$. Equating $(\Delta t)_1$ to the lifetime of the star $t_*$ and making
use of Eq.~(\ref{eq:t0}) leads to the estimate (\ref{eq:mtrans}).

\bibliography{mainbibv1}

\begin{thebibliography}{55}
\expandafter\ifx\csname natexlab\endcsname\relax\def\natexlab#1{#1}\fi
\expandafter\ifx\csname bibnamefont\endcsname\relax
  \def\bibnamefont#1{#1}\fi
\expandafter\ifx\csname bibfnamefont\endcsname\relax
  \def\bibfnamefont#1{#1}\fi
\expandafter\ifx\csname citenamefont\endcsname\relax
  \def\citenamefont#1{#1}\fi
\expandafter\ifx\csname url\endcsname\relax
  \def\url#1{\texttt{#1}}\fi
\expandafter\ifx\csname urlprefix\endcsname\relax\def\urlprefix{URL }\fi
\providecommand{\bibinfo}[2]{#2}
\providecommand{\eprint}[2][]{\url{#2}}

\bibitem[{\citenamefont{Bertone et~al.}(2005)\citenamefont{Bertone, Hooper, and
  Silk}}]{Bertone:2004pz}
\bibinfo{author}{\bibfnamefont{G.}~\bibnamefont{Bertone}},
  \bibinfo{author}{\bibfnamefont{D.}~\bibnamefont{Hooper}}, \bibnamefont{and}
  \bibinfo{author}{\bibfnamefont{J.}~\bibnamefont{Silk}},
  \bibinfo{journal}{Phys.Rept.} \textbf{\bibinfo{volume}{405}},
  \bibinfo{pages}{279} (\bibinfo{year}{2005}), \eprint{hep-ph/0404175}.

\bibitem[{\citenamefont{Bergstrom}(2012)}]{Bergstrom:2012fi}
\bibinfo{author}{\bibfnamefont{L.}~\bibnamefont{Bergstrom}}
  (\bibinfo{year}{2012}), \eprint{1205.4882}.

\bibitem[{\citenamefont{Komatsu et~al.}(2011)}]{Komatsu:2010fb}
\bibinfo{author}{\bibfnamefont{E.}~\bibnamefont{Komatsu}} \bibnamefont{et~al.}
  (\bibinfo{collaboration}{WMAP Collaboration}),
  \bibinfo{journal}{Astrophys.J.Suppl.} \textbf{\bibinfo{volume}{192}},
  \bibinfo{pages}{18} (\bibinfo{year}{2011}), \eprint{1001.4538}.

\bibitem[{\citenamefont{Hawking}(1971)}]{Hawking:1971ei}
\bibinfo{author}{\bibfnamefont{S.}~\bibnamefont{Hawking}},
  \bibinfo{journal}{Mon.Not.Roy.Astron.Soc.} \textbf{\bibinfo{volume}{152}},
  \bibinfo{pages}{75} (\bibinfo{year}{1971}).

\bibitem[{\citenamefont{Dolgov and Silk}(1993)}]{Dolgov:1992pu}
\bibinfo{author}{\bibfnamefont{A.}~\bibnamefont{Dolgov}} \bibnamefont{and}
  \bibinfo{author}{\bibfnamefont{J.}~\bibnamefont{Silk}},
  \bibinfo{journal}{Phys.Rev.} \textbf{\bibinfo{volume}{D47}},
  \bibinfo{pages}{4244} (\bibinfo{year}{1993}).

\bibitem[{\citenamefont{Green and Liddle}(1999)}]{Green:1999xm}
\bibinfo{author}{\bibfnamefont{A.~M.} \bibnamefont{Green}} \bibnamefont{and}
  \bibinfo{author}{\bibfnamefont{A.~R.} \bibnamefont{Liddle}},
  \bibinfo{journal}{Phys.Rev.} \textbf{\bibinfo{volume}{D60}},
  \bibinfo{pages}{063509} (\bibinfo{year}{1999}), \eprint{astro-ph/9901268}.

\bibitem[{\citenamefont{Hawking}(1974)}]{Hawking:1974rv}
\bibinfo{author}{\bibfnamefont{S.}~\bibnamefont{Hawking}},
  \bibinfo{journal}{Nature} \textbf{\bibinfo{volume}{248}}, \bibinfo{pages}{30}
  (\bibinfo{year}{1974}).

\bibitem[{\citenamefont{Page and Hawking}(1976)}]{Page:1976wx}
\bibinfo{author}{\bibfnamefont{D.~N.} \bibnamefont{Page}} \bibnamefont{and}
  \bibinfo{author}{\bibfnamefont{S.}~\bibnamefont{Hawking}},
  \bibinfo{journal}{Astrophys.J.} \textbf{\bibinfo{volume}{206}},
  \bibinfo{pages}{1} (\bibinfo{year}{1976}).

\bibitem[{\citenamefont{Sreekumar et~al.}(1998)}]{Sreekumar:1997un}
\bibinfo{author}{\bibfnamefont{P.}~\bibnamefont{Sreekumar}}
  \bibnamefont{et~al.} (\bibinfo{collaboration}{EGRET Collaboration}),
  \bibinfo{journal}{Astrophys.J.} \textbf{\bibinfo{volume}{494}},
  \bibinfo{pages}{523} (\bibinfo{year}{1998}), \eprint{astro-ph/9709257}.

\bibitem[{\citenamefont{Carr et~al.}(2010)\citenamefont{Carr, Kohri, Sendouda,
  and Yokoyama}}]{Carr:2009jm}
\bibinfo{author}{\bibfnamefont{B.}~\bibnamefont{Carr}},
  \bibinfo{author}{\bibfnamefont{K.}~\bibnamefont{Kohri}},
  \bibinfo{author}{\bibfnamefont{Y.}~\bibnamefont{Sendouda}}, \bibnamefont{and}
  \bibinfo{author}{\bibfnamefont{J.}~\bibnamefont{Yokoyama}},
  \bibinfo{journal}{Phys.Rev.} \textbf{\bibinfo{volume}{D81}},
  \bibinfo{pages}{104019} (\bibinfo{year}{2010}), \eprint{astro-ph/0912.5297}.

\bibitem[{\citenamefont{{Gould}}(1992)}]{Gould:1992wz}
\bibinfo{author}{\bibfnamefont{A.}~\bibnamefont{{Gould}}},
  \bibinfo{journal}{\apjl} \textbf{\bibinfo{volume}{386}}, \bibinfo{pages}{L5}
  (\bibinfo{year}{1992}).

\bibitem[{\citenamefont{{Barnacka} et~al.}(2012)\citenamefont{{Barnacka},
  {Glicenstein}, and {Moderski}}}]{Barnacka:2012bm}
\bibinfo{author}{\bibfnamefont{A.}~\bibnamefont{{Barnacka}}},
  \bibinfo{author}{\bibfnamefont{J.-F.} \bibnamefont{{Glicenstein}}},
  \bibnamefont{and}
  \bibinfo{author}{\bibfnamefont{R.}~\bibnamefont{{Moderski}}},
  \bibinfo{journal}{\prd} \textbf{\bibinfo{volume}{86}}, \bibinfo{eid}{043001}
  (\bibinfo{year}{2012}), \eprint{1204.2056}.

\bibitem[{\citenamefont{Tisserand et~al.}(2007)}]{Tisserand:2006zx}
\bibinfo{author}{\bibfnamefont{P.}~\bibnamefont{Tisserand}}
  \bibnamefont{et~al.} (\bibinfo{collaboration}{EROS-2 Collaboration}),
  \bibinfo{journal}{Astron.Astrophys.} \textbf{\bibinfo{volume}{469}},
  \bibinfo{pages}{387} (\bibinfo{year}{2007}), \eprint{astro-ph/0607207}.

\bibitem[{\citenamefont{Ricotti et~al.}(2007)\citenamefont{Ricotti, Ostriker,
  and Mack}}]{Ricotti:2007au}
\bibinfo{author}{\bibfnamefont{M.}~\bibnamefont{Ricotti}},
  \bibinfo{author}{\bibfnamefont{J.~P.} \bibnamefont{Ostriker}},
  \bibnamefont{and} \bibinfo{author}{\bibfnamefont{K.~J.} \bibnamefont{Mack}}
  (\bibinfo{year}{2007}), \eprint{0709.0524}.

\bibitem[{\citenamefont{Kouvaris and
  Tinyakov}(2011{\natexlab{a}})}]{Kouvaris:2011fi}
\bibinfo{author}{\bibfnamefont{C.}~\bibnamefont{Kouvaris}} \bibnamefont{and}
  \bibinfo{author}{\bibfnamefont{P.}~\bibnamefont{Tinyakov}},
  \bibinfo{journal}{Phys.Rev.Lett.} \textbf{\bibinfo{volume}{107}},
  \bibinfo{pages}{091301} (\bibinfo{year}{2011}{\natexlab{a}}),
  \eprint{1104.0382}.

\bibitem[{\citenamefont{Kouvaris}(2012)}]{Kouvaris:2011gb}
\bibinfo{author}{\bibfnamefont{C.}~\bibnamefont{Kouvaris}},
  \bibinfo{journal}{Phys.Rev.Lett.} \textbf{\bibinfo{volume}{108}},
  \bibinfo{pages}{191301} (\bibinfo{year}{2012}), \eprint{1111.4364}.

\bibitem[{\citenamefont{Kouvaris and
  Tinyakov}(2011{\natexlab{b}})}]{Kouvaris:2010jy}
\bibinfo{author}{\bibfnamefont{C.}~\bibnamefont{Kouvaris}} \bibnamefont{and}
  \bibinfo{author}{\bibfnamefont{P.}~\bibnamefont{Tinyakov}},
  \bibinfo{journal}{Phys.Rev.} \textbf{\bibinfo{volume}{D83}},
  \bibinfo{pages}{083512} (\bibinfo{year}{2011}{\natexlab{b}}),
  \eprint{1012.2039}.

\bibitem[{\citenamefont{{Mouschovias} and
  {Morton}}(1991)}]{1991ApJ...371..296M}
\bibinfo{author}{\bibfnamefont{T.~C.} \bibnamefont{{Mouschovias}}}
  \bibnamefont{and} \bibinfo{author}{\bibfnamefont{S.~A.}
  \bibnamefont{{Morton}}}, \bibinfo{journal}{\apj}
  \textbf{\bibinfo{volume}{371}}, \bibinfo{pages}{296} (\bibinfo{year}{1991}).

\bibitem[{\citenamefont{{Shu} et~al.}(1987)\citenamefont{{Shu}, {Adams}, and
  {Lizano}}}]{1987ARA&amp;A..25...23S}
\bibinfo{author}{\bibfnamefont{F.~H.} \bibnamefont{{Shu}}},
  \bibinfo{author}{\bibfnamefont{F.~C.} \bibnamefont{{Adams}}},
  \bibnamefont{and} \bibinfo{author}{\bibfnamefont{S.}~\bibnamefont{{Lizano}}},
  \bibinfo{journal}{\araa} \textbf{\bibinfo{volume}{25}}, \bibinfo{pages}{23}
  (\bibinfo{year}{1987}).

\bibitem[{\citenamefont{{Larson}}(1969)}]{Larson69}
\bibinfo{author}{\bibfnamefont{R.~B.} \bibnamefont{{Larson}}},
  \bibinfo{journal}{\mnras} \textbf{\bibinfo{volume}{145}},
  \bibinfo{pages}{271} (\bibinfo{year}{1969}).

\bibitem[{\citenamefont{{Boss} and {Yorke}}(1995)}]{Boss95}
\bibinfo{author}{\bibfnamefont{A.~P.} \bibnamefont{{Boss}}} \bibnamefont{and}
  \bibinfo{author}{\bibfnamefont{H.~W.} \bibnamefont{{Yorke}}},
  \bibinfo{journal}{\apjl} \textbf{\bibinfo{volume}{439}}, \bibinfo{pages}{L55}
  (\bibinfo{year}{1995}).

\bibitem[{\citenamefont{{Bate}}(1998)}]{Bate:1998hg}
\bibinfo{author}{\bibfnamefont{M.~R.} \bibnamefont{{Bate}}},
  \bibinfo{journal}{\apjl} \textbf{\bibinfo{volume}{508}}, \bibinfo{pages}{L95}
  (\bibinfo{year}{1998}), \eprint{astro-ph/9810397}.

\bibitem[{\citenamefont{Blumenthal et~al.}(1986)\citenamefont{Blumenthal,
  Faber, Flores, and Primack}}]{Blumenthal:1985qy}
\bibinfo{author}{\bibfnamefont{G.~R.} \bibnamefont{Blumenthal}},
  \bibinfo{author}{\bibfnamefont{S.}~\bibnamefont{Faber}},
  \bibinfo{author}{\bibfnamefont{R.}~\bibnamefont{Flores}}, \bibnamefont{and}
  \bibinfo{author}{\bibfnamefont{J.~R.} \bibnamefont{Primack}},
  \bibinfo{journal}{Astrophys.J.} \textbf{\bibinfo{volume}{301}},
  \bibinfo{pages}{27} (\bibinfo{year}{1986}).

\bibitem[{\citenamefont{Gnedin et~al.}(2004)\citenamefont{Gnedin, Kravtsov,
  Klypin, and Nagai}}]{Gnedin:2004cx}
\bibinfo{author}{\bibfnamefont{O.~Y.} \bibnamefont{Gnedin}},
  \bibinfo{author}{\bibfnamefont{A.~V.} \bibnamefont{Kravtsov}},
  \bibinfo{author}{\bibfnamefont{A.~A.} \bibnamefont{Klypin}},
  \bibnamefont{and} \bibinfo{author}{\bibfnamefont{D.}~\bibnamefont{Nagai}},
  \bibinfo{journal}{Astrophys.J.} \textbf{\bibinfo{volume}{616}},
  \bibinfo{pages}{16} (\bibinfo{year}{2004}), \eprint{astro-ph/0406247}.

\bibitem[{\citenamefont{Sellwood and McGaugh}(2005)}]{Sellwood:2005pq}
\bibinfo{author}{\bibfnamefont{J.~A.} \bibnamefont{Sellwood}} \bibnamefont{and}
  \bibinfo{author}{\bibfnamefont{S.~S.} \bibnamefont{McGaugh}},
  \bibinfo{journal}{Astrophys.J.} \textbf{\bibinfo{volume}{634}},
  \bibinfo{pages}{70} (\bibinfo{year}{2005}), \eprint{astro-ph/0507589}.

\bibitem[{\citenamefont{{Derishev} and
  {Belyanin}}(1999)}]{1999A&amp;A...343....1D}
\bibinfo{author}{\bibfnamefont{E.~V.} \bibnamefont{{Derishev}}}
  \bibnamefont{and} \bibinfo{author}{\bibfnamefont{A.~A.}
  \bibnamefont{{Belyanin}}}, \bibinfo{journal}{\aap}
  \textbf{\bibinfo{volume}{343}}, \bibinfo{pages}{1} (\bibinfo{year}{1999}).

\bibitem[{\citenamefont{Tremaine and Gunn}(1979)}]{Tremaine:1979we}
\bibinfo{author}{\bibfnamefont{S.}~\bibnamefont{Tremaine}} \bibnamefont{and}
  \bibinfo{author}{\bibfnamefont{J.}~\bibnamefont{Gunn}},
  \bibinfo{journal}{Phys.Rev.Lett.} \textbf{\bibinfo{volume}{42}},
  \bibinfo{pages}{407} (\bibinfo{year}{1979}).

\bibitem[{\citenamefont{{Steigman} et~al.}(1978)\citenamefont{{Steigman},
  {Sarazin}, {Quintana}, and {Faulkner}}}]{1978AJ.....83.1050S}
\bibinfo{author}{\bibfnamefont{G.}~\bibnamefont{{Steigman}}},
  \bibinfo{author}{\bibfnamefont{C.~L.} \bibnamefont{{Sarazin}}},
  \bibinfo{author}{\bibfnamefont{H.}~\bibnamefont{{Quintana}}},
  \bibnamefont{and}
  \bibinfo{author}{\bibfnamefont{J.}~\bibnamefont{{Faulkner}}},
  \bibinfo{journal}{\aj} \textbf{\bibinfo{volume}{83}}, \bibinfo{pages}{1050}
  (\bibinfo{year}{1978}).

\bibitem[{\citenamefont{Spolyar et~al.}(2008)\citenamefont{Spolyar, Freese, and
  Gondolo}}]{Spolyar:2007qv}
\bibinfo{author}{\bibfnamefont{D.}~\bibnamefont{Spolyar}},
  \bibinfo{author}{\bibfnamefont{K.}~\bibnamefont{Freese}}, \bibnamefont{and}
  \bibinfo{author}{\bibfnamefont{P.}~\bibnamefont{Gondolo}},
  \bibinfo{journal}{Phys.Rev.Lett.} \textbf{\bibinfo{volume}{100}},
  \bibinfo{pages}{051101} (\bibinfo{year}{2008}), \eprint{astro-ph/0705.0521}.

\bibitem[{\citenamefont{Bambi et~al.}(2009)\citenamefont{Bambi, Spolyar,
  Dolgov, Freese, and Volonteri}}]{Bambi:2008uc}
\bibinfo{author}{\bibfnamefont{C.}~\bibnamefont{Bambi}},
  \bibinfo{author}{\bibfnamefont{D.}~\bibnamefont{Spolyar}},
  \bibinfo{author}{\bibfnamefont{A.~D.} \bibnamefont{Dolgov}},
  \bibinfo{author}{\bibfnamefont{K.}~\bibnamefont{Freese}}, \bibnamefont{and}
  \bibinfo{author}{\bibfnamefont{M.}~\bibnamefont{Volonteri}},
  \bibinfo{journal}{Mon.Not.Roy.Astron.Soc.} \textbf{\bibinfo{volume}{399}},
  \bibinfo{pages}{1347} (\bibinfo{year}{2009}), \eprint{astro-ph/0812.0585}.

\bibitem[{\citenamefont{{Jesseit} et~al.}(2002)\citenamefont{{Jesseit}, {Naab},
  and {Burkert}}}]{Jesseit:2002tj}
\bibinfo{author}{\bibfnamefont{R.}~\bibnamefont{{Jesseit}}},
  \bibinfo{author}{\bibfnamefont{T.}~\bibnamefont{{Naab}}}, \bibnamefont{and}
  \bibinfo{author}{\bibfnamefont{A.}~\bibnamefont{{Burkert}}},
  \bibinfo{journal}{\apjl} \textbf{\bibinfo{volume}{571}}, \bibinfo{pages}{L89}
  (\bibinfo{year}{2002}), \eprint{astro-ph/0204164}.

\bibitem[{\citenamefont{{Dotter} et~al.}(2010)\citenamefont{{Dotter},
  {Sarajedini}, {Anderson}, {Aparicio}, {Bedin}, {Chaboyer}, {Majewski},
  {Mar{\'{\i}}n-Franch}, {Milone}, {Paust} et~al.}}]{Dotter2010}
\bibinfo{author}{\bibfnamefont{A.}~\bibnamefont{{Dotter}}},
  \bibinfo{author}{\bibfnamefont{A.}~\bibnamefont{{Sarajedini}}},
  \bibinfo{author}{\bibfnamefont{J.}~\bibnamefont{{Anderson}}},
  \bibinfo{author}{\bibfnamefont{A.}~\bibnamefont{{Aparicio}}},
  \bibinfo{author}{\bibfnamefont{L.~R.} \bibnamefont{{Bedin}}},
  \bibinfo{author}{\bibfnamefont{B.}~\bibnamefont{{Chaboyer}}},
  \bibinfo{author}{\bibfnamefont{S.}~\bibnamefont{{Majewski}}},
  \bibinfo{author}{\bibfnamefont{A.}~\bibnamefont{{Mar{\'{\i}}n-Franch}}},
  \bibinfo{author}{\bibfnamefont{A.}~\bibnamefont{{Milone}}},
  \bibinfo{author}{\bibfnamefont{N.}~\bibnamefont{{Paust}}},
  \bibnamefont{et~al.}, \bibinfo{journal}{\apj} \textbf{\bibinfo{volume}{708}},
  \bibinfo{pages}{698} (\bibinfo{year}{2010}), \eprint{0911.2469}.

\bibitem[{\citenamefont{{Peebles}}(1984)}]{Peebles1984}
\bibinfo{author}{\bibfnamefont{P.~J.~E.} \bibnamefont{{Peebles}}},
  \bibinfo{journal}{\apj} \textbf{\bibinfo{volume}{277}}, \bibinfo{pages}{470}
  (\bibinfo{year}{1984}).

\bibitem[{\citenamefont{{Bromm} and {Clarke}}(2002)}]{Bromm2002}
\bibinfo{author}{\bibfnamefont{V.}~\bibnamefont{{Bromm}}} \bibnamefont{and}
  \bibinfo{author}{\bibfnamefont{C.~J.} \bibnamefont{{Clarke}}},
  \bibinfo{journal}{\apjl} \textbf{\bibinfo{volume}{566}}, \bibinfo{pages}{L1}
  (\bibinfo{year}{2002}), \eprint{arXiv:astro-ph/0201066}.

\bibitem[{\citenamefont{{Mashchenko} and
  {Sills}}(2005{\natexlab{a}})}]{MS2005a}
\bibinfo{author}{\bibfnamefont{S.}~\bibnamefont{{Mashchenko}}}
  \bibnamefont{and} \bibinfo{author}{\bibfnamefont{A.}~\bibnamefont{{Sills}}},
  \bibinfo{journal}{\apj} \textbf{\bibinfo{volume}{619}}, \bibinfo{pages}{243}
  (\bibinfo{year}{2005}{\natexlab{a}}), \eprint{arXiv:astro-ph/0409605}.

\bibitem[{\citenamefont{{Moore} et~al.}(2006)\citenamefont{{Moore}, {Diemand},
  {Madau}, {Zemp}, and {Stadel}}}]{Moore2006}
\bibinfo{author}{\bibfnamefont{B.}~\bibnamefont{{Moore}}},
  \bibinfo{author}{\bibfnamefont{J.}~\bibnamefont{{Diemand}}},
  \bibinfo{author}{\bibfnamefont{P.}~\bibnamefont{{Madau}}},
  \bibinfo{author}{\bibfnamefont{M.}~\bibnamefont{{Zemp}}}, \bibnamefont{and}
  \bibinfo{author}{\bibfnamefont{J.}~\bibnamefont{{Stadel}}},
  \bibinfo{journal}{\mnras} \textbf{\bibinfo{volume}{368}},
  \bibinfo{pages}{563} (\bibinfo{year}{2006}), \eprint{arXiv:astro-ph/0510370}.

\bibitem[{\citenamefont{{Boley} et~al.}(2009)\citenamefont{{Boley}, {Lake},
  {Read}, and {Teyssier}}}]{Boley2009}
\bibinfo{author}{\bibfnamefont{A.~C.} \bibnamefont{{Boley}}},
  \bibinfo{author}{\bibfnamefont{G.}~\bibnamefont{{Lake}}},
  \bibinfo{author}{\bibfnamefont{J.}~\bibnamefont{{Read}}}, \bibnamefont{and}
  \bibinfo{author}{\bibfnamefont{R.}~\bibnamefont{{Teyssier}}},
  \bibinfo{journal}{\apjl} \textbf{\bibinfo{volume}{706}},
  \bibinfo{pages}{L192} (\bibinfo{year}{2009}), \eprint{0908.1254}.

\bibitem[{\citenamefont{{Griffen} et~al.}(2010)\citenamefont{{Griffen},
  {Drinkwater}, {Thomas}, {Helly}, and {Pimbblet}}}]{Griffen2010}
\bibinfo{author}{\bibfnamefont{B.~F.} \bibnamefont{{Griffen}}},
  \bibinfo{author}{\bibfnamefont{M.~J.} \bibnamefont{{Drinkwater}}},
  \bibinfo{author}{\bibfnamefont{P.~A.} \bibnamefont{{Thomas}}},
  \bibinfo{author}{\bibfnamefont{J.~C.} \bibnamefont{{Helly}}},
  \bibnamefont{and} \bibinfo{author}{\bibfnamefont{K.~A.}
  \bibnamefont{{Pimbblet}}}, \bibinfo{journal}{\mnras}
  \textbf{\bibinfo{volume}{405}}, \bibinfo{pages}{375} (\bibinfo{year}{2010}),
  \eprint{0910.0310}.

\bibitem[{\citenamefont{{Fall} and {Rees}}(1985)}]{Fall1985}
\bibinfo{author}{\bibfnamefont{S.~M.} \bibnamefont{{Fall}}} \bibnamefont{and}
  \bibinfo{author}{\bibfnamefont{M.~J.} \bibnamefont{{Rees}}},
  \bibinfo{journal}{\apj} \textbf{\bibinfo{volume}{298}}, \bibinfo{pages}{18}
  (\bibinfo{year}{1985}).

\bibitem[{\citenamefont{{Ashman} and {Zepf}}(1992)}]{Ashman1992}
\bibinfo{author}{\bibfnamefont{K.~M.} \bibnamefont{{Ashman}}} \bibnamefont{and}
  \bibinfo{author}{\bibfnamefont{S.~E.} \bibnamefont{{Zepf}}},
  \bibinfo{journal}{\apj} \textbf{\bibinfo{volume}{384}}, \bibinfo{pages}{50}
  (\bibinfo{year}{1992}).

\bibitem[{\citenamefont{{Kravtsov} and {Gnedin}}(2005)}]{Kravtsov2005}
\bibinfo{author}{\bibfnamefont{A.~V.} \bibnamefont{{Kravtsov}}}
  \bibnamefont{and} \bibinfo{author}{\bibfnamefont{O.~Y.}
  \bibnamefont{{Gnedin}}}, \bibinfo{journal}{\apj}
  \textbf{\bibinfo{volume}{623}}, \bibinfo{pages}{650} (\bibinfo{year}{2005}),
  \eprint{arXiv:astro-ph/0305199}.

\bibitem[{\citenamefont{{Muratov} and {Gnedin}}(2010)}]{Muratov2010}
\bibinfo{author}{\bibfnamefont{A.~L.} \bibnamefont{{Muratov}}}
  \bibnamefont{and} \bibinfo{author}{\bibfnamefont{O.~Y.}
  \bibnamefont{{Gnedin}}}, \bibinfo{journal}{\apj}
  \textbf{\bibinfo{volume}{718}}, \bibinfo{pages}{1266} (\bibinfo{year}{2010}),
  \eprint{1002.1325}.

\bibitem[{\citenamefont{{Brodie} and {Strader}}(2006)}]{Brodie2006}
\bibinfo{author}{\bibfnamefont{J.~P.} \bibnamefont{{Brodie}}} \bibnamefont{and}
  \bibinfo{author}{\bibfnamefont{J.}~\bibnamefont{{Strader}}},
  \bibinfo{journal}{\araa} \textbf{\bibinfo{volume}{44}}, \bibinfo{pages}{193}
  (\bibinfo{year}{2006}), \eprint{arXiv:astro-ph/0602601}.

\bibitem[{\citenamefont{{Moore}}(1996)}]{Moore1996}
\bibinfo{author}{\bibfnamefont{B.}~\bibnamefont{{Moore}}},
  \bibinfo{journal}{\apjl} \textbf{\bibinfo{volume}{461}}, \bibinfo{pages}{L13}
  (\bibinfo{year}{1996}), \eprint{arXiv:astro-ph/9511147}.

\bibitem[{\citenamefont{{Mashchenko} and
  {Sills}}(2005{\natexlab{b}})}]{MS2005b}
\bibinfo{author}{\bibfnamefont{S.}~\bibnamefont{{Mashchenko}}}
  \bibnamefont{and} \bibinfo{author}{\bibfnamefont{A.}~\bibnamefont{{Sills}}},
  \bibinfo{journal}{\apj} \textbf{\bibinfo{volume}{619}}, \bibinfo{pages}{258}
  (\bibinfo{year}{2005}{\natexlab{b}}), \eprint{arXiv:astro-ph/0409606}.

\bibitem[{\citenamefont{Bertone and Fairbairn}(2008)}]{Bertone:2007ae}
\bibinfo{author}{\bibfnamefont{G.}~\bibnamefont{Bertone}} \bibnamefont{and}
  \bibinfo{author}{\bibfnamefont{M.}~\bibnamefont{Fairbairn}},
  \bibinfo{journal}{Phys.Rev.} \textbf{\bibinfo{volume}{D77}},
  \bibinfo{pages}{043515} (\bibinfo{year}{2008}), \eprint{0709.1485}.

\bibitem[{\citenamefont{{Harris}}(1996)}]{GC-velocity-dispersion}
\bibinfo{author}{\bibfnamefont{W.~E.} \bibnamefont{{Harris}}},
  \bibinfo{journal}{\aj} \textbf{\bibinfo{volume}{112}}, \bibinfo{pages}{1487}
  (\bibinfo{year}{1996}).

\bibitem[{\citenamefont{Kirk et~al.}(2005)\citenamefont{Kirk, Ward-Thompson,
  and Andre}}]{Kirk:2005ng}
\bibinfo{author}{\bibfnamefont{J.~M.} \bibnamefont{Kirk}},
  \bibinfo{author}{\bibfnamefont{D.}~\bibnamefont{Ward-Thompson}},
  \bibnamefont{and} \bibinfo{author}{\bibfnamefont{P.}~\bibnamefont{Andre}},
  \bibinfo{journal}{Mon.Not.Roy.Astron.Soc.} \textbf{\bibinfo{volume}{360}},
  \bibinfo{pages}{1506} (\bibinfo{year}{2005}), \eprint{astro-ph/0505190}.

\bibitem[{\citenamefont{{Baumgardt} et~al.}(2009)\citenamefont{{Baumgardt},
  {C{\^o}t{\'e}}, {Hilker}, {Rejkuba}, {Mieske}, {Djorgovski}, and
  {Stetson}}}]{Baumgardt2009}
\bibinfo{author}{\bibfnamefont{H.}~\bibnamefont{{Baumgardt}}},
  \bibinfo{author}{\bibfnamefont{P.}~\bibnamefont{{C{\^o}t{\'e}}}},
  \bibinfo{author}{\bibfnamefont{M.}~\bibnamefont{{Hilker}}},
  \bibinfo{author}{\bibfnamefont{M.}~\bibnamefont{{Rejkuba}}},
  \bibinfo{author}{\bibfnamefont{S.}~\bibnamefont{{Mieske}}},
  \bibinfo{author}{\bibfnamefont{S.~G.} \bibnamefont{{Djorgovski}}},
  \bibnamefont{and}
  \bibinfo{author}{\bibfnamefont{P.}~\bibnamefont{{Stetson}}},
  \bibinfo{journal}{\mnras} \textbf{\bibinfo{volume}{396}},
  \bibinfo{pages}{2051} (\bibinfo{year}{2009}), \eprint{0904.3329}.

\bibitem[{\citenamefont{Conroy et~al.}(2011)\citenamefont{Conroy, Loeb, and
  Spergel}}]{Conroy:2010bs}
\bibinfo{author}{\bibfnamefont{C.}~\bibnamefont{Conroy}},
  \bibinfo{author}{\bibfnamefont{A.}~\bibnamefont{Loeb}}, \bibnamefont{and}
  \bibinfo{author}{\bibfnamefont{D.}~\bibnamefont{Spergel}},
  \bibinfo{journal}{Astrophys.J.} \textbf{\bibinfo{volume}{741}},
  \bibinfo{pages}{72} (\bibinfo{year}{2011}), \eprint{1010.5783}.

\bibitem[{\citenamefont{{Mackey} and {van den Bergh}}(2005)}]{Mackey2005}
\bibinfo{author}{\bibfnamefont{A.~D.} \bibnamefont{{Mackey}}} \bibnamefont{and}
  \bibinfo{author}{\bibfnamefont{S.}~\bibnamefont{{van den Bergh}}},
  \bibinfo{journal}{\mnras} \textbf{\bibinfo{volume}{360}},
  \bibinfo{pages}{631} (\bibinfo{year}{2005}), \eprint{arXiv:astro-ph/0504142}.

\bibitem[{\citenamefont{{Cohen} and {Kirby}}(2012)}]{Cohen2012}
\bibinfo{author}{\bibfnamefont{J.~G.} \bibnamefont{{Cohen}}} \bibnamefont{and}
  \bibinfo{author}{\bibfnamefont{E.~N.} \bibnamefont{{Kirby}}},
  \bibinfo{journal}{ArXiv e-prints}  (\bibinfo{year}{2012}),
  \eprint{1209.2705}.

\bibitem[{\citenamefont{{Cohen} et~al.}(2010)\citenamefont{{Cohen}, {Kirby},
  {Simon}, and {Geha}}}]{Cohen2010}
\bibinfo{author}{\bibfnamefont{J.~G.} \bibnamefont{{Cohen}}},
  \bibinfo{author}{\bibfnamefont{E.~N.} \bibnamefont{{Kirby}}},
  \bibinfo{author}{\bibfnamefont{J.~D.} \bibnamefont{{Simon}}},
  \bibnamefont{and} \bibinfo{author}{\bibfnamefont{M.}~\bibnamefont{{Geha}}},
  \bibinfo{journal}{\apj} \textbf{\bibinfo{volume}{725}}, \bibinfo{pages}{288}
  (\bibinfo{year}{2010}), \eprint{1010.0031}.

\bibitem[{\citenamefont{{Chandrasekhar}}(1949)}]{1949RvMP...21..383C}
\bibinfo{author}{\bibfnamefont{S.}~\bibnamefont{{Chandrasekhar}}},
  \bibinfo{journal}{Reviews of Modern Physics} \textbf{\bibinfo{volume}{21}},
  \bibinfo{pages}{383} (\bibinfo{year}{1949}).

\bibitem[{\citenamefont{{Binney} and {Tremaine}}(1987)}]{1987gady.book.....B}
\bibinfo{author}{\bibfnamefont{J.}~\bibnamefont{{Binney}}} \bibnamefont{and}
  \bibinfo{author}{\bibfnamefont{S.}~\bibnamefont{{Tremaine}}},
  \emph{\bibinfo{title}{{Galactic dynamics}}} (\bibinfo{year}{1987}).

\end{thebibliography}
\bibliographystyle{apsrev}

\end{document}